\documentclass[article,prb,twocolumn,aps,showpacs,amsmath,amssymb]{revtex4-1}
\usepackage{graphicx}

\begin{document}

\title{Effects of composition and chemical disorder on the magnetocrystalline anisotropy of Fe$_x$Pt$_{1-x}$ alloys}

\author{C.J.~Aas$^1$}
\author{L.~Szunyogh$^{2}$}
\author{R.W.~Chantrell$^1$}
\affiliation{$^1$ Department of Physics, University of York, York YO10 5DD, United Kingdom}
\affiliation{$^2$ Department of Theoretical Physics and Condensed Matter Research Group of Hungarian Academy of Sciences, Budapest University of Technology and Economics, Budafoki \'ut 8. H1111 Budapest, Hungary}

\date{\today}

\begin{abstract}

We perform first principles calculations of the magnetocrystalline anisotropy energy (MAE) of the L1$_0$-like Fe$_x$Pt$_{1-x}$ samples studied experimentally by Barmak and co-workers in [J.~Appl.~Phys.~\textbf{98} (2005) 033904].  The variation of composition and long-range chemical order in the samples was studied in terms of the coherent potential approximation. In accordance with experimental observations, we find that, in the presence of long-range chemical disorder, Fe-rich samples exhibit a larger MAE than stoichiometric FePt.  By considering the site- and species-resolved contributions to the MAE, we infer that the MAE is primarily a function of the degree of completeness of the nominal Fe layers in the L1$_0$ FePt structure.
\end{abstract}


\maketitle
Due to its extraordinarily high magnetocrystalline anisotropy energy (MAE), L1$_0$ FePt is of considerable interest to the development of ultrahigh density magnetic recording applications, in particular, for heat-assisted magnetic recording (HAMR).  The L1$_0$ phase of Fe$_{50}$Pt$_{50}$ is a layered face-centered tetragonal structure, exhibiting alternating Fe and Pt layers along the $(0 0 1)$ direction.  FePt also exhibits stable FePt$_{3}$ and Fe$_{3}$Pt phases as well as a chemically disordered, cubic phase.\cite{whang, massalski}  Accordingly, FePt exhibits phase transitions with respect to composition as well as to chemical order and understanding the related effects on the magnetic properties is an important issue.  The large effect of chemical disorder on the MAE of Fe$_{50}$Pt$_{50}$ has already been outlined both experimentally\cite{disorder1,expl,barmak} and theoretically.\cite{disorder2,apl-letter}\\*

The degree of long-range chemical order is quantified in terms of a chemical order parameter.\cite{shockley, warren}
The L1$_0$ Fe$_x$Pt$_{1-x}$ alloy is modeled by a repeating sequence of two atomic layers, characterized by compositions Fe$_{r_{Fe}}$Pt$_{1-r_{Fe}}$ and Fe$_{1-r_{Pt}}$Pt$_{r_{Pt}}$, respectively. The fractions, $r_{Fe}$ and $r_{Pt}$, are related to each other through the condition, $1+r_{Fe}-r_{Pt}=2x$. Furthermore, set by the requirement,  $r_{Fe} \ge 1-r_{Pt}$ (the case of $r_{Fe} < 1-r_{Pt}$ can simply be obtained by interchanging the two types of layers), the range of $r_{Fe}$ is confined to  $r_{Fe} \ge x$  ($r_{Pt} \ge 1-x$), whereby obviously  $r_{Fe} \le {\rm min}(1,2x)$ ($r_{Pt} \le {\rm min}(1,2-2x)$). The chemical order parameter $s$ is then defined by
\begin{equation}
 s=2 (r_{Fe}-x) =2(r_{Pt}-1+x) \: ,
 \label{eq:chemord}
\end{equation}
and ranges from $0$ to max($2-2x,2x$). Denoting the compositions of the two repeating layers as $(A,B)$, the case of complete disorder refers to  the
compositions (Fe$_x$Pt$_{1-x}$, Fe$_x$Pt$_{1-x}$) and the maximum order to (Fe, Fe$_{2x-1}$Pt$_{2-2x}$)
for $x \ge 0.5$ and to (Fe$_{2x}$Pt$_{1-2x}$,Pt) for $x \le 0.5$. Note that only in case of $x=0.5$ can the order
parameter reach the value $s=1$. In the following, we refer to the two layers as the nominal Fe layer and the nominal Pt layer, respectively.\\*

\begin{table}[htp]\centering
\caption{Summary of the experimental data obtained for four samples of Fe$_x$Pt$_{1-x}$  in Ref.~\onlinecite{barmak}.\label{table:samplesBarmak}}
\begin{tabular}{@{}ccccccc@{}}\toprule
\textbf{Sample} & $x$  (\%) & a (\AA) & c (\AA) & c/a & $s$  & $K$ (meV/atom)\\ \colrule

1 & 46.2 & 3.870 & 3.721 & 0.961 & 0.89 & 0.453 \\
2 & 51.1 & 3.863 & 3.710 & 0.960 & 0.93 & 0.709 \\
3 & 52.0 & 3.857 & 3.706 & 0.961 & 0.89 & 0.775 \\
4 & 55.4 & 3.839 & 3.704 & 0.965 & 0.72 & N/A \\
 \botrule
\end{tabular}
\end{table}

Our present study was motivated by the work of Barmak and co-workers,\cite{barmak} who investigated the MAE of four Fe$_x$Pt$_{1-x}$ samples differing in composition and degree of chemical order.
Table \ref{table:samplesBarmak} summarizes the experimental geometrical and compositional data,
as well as the measured MAE values for the FePt samples studied in Ref.~\onlinecite{barmak}.
Note that for sample no. 4 the MAE could not be determined.
One of the main conclusions of Ref.~\onlinecite{barmak} is that slightly Fe-rich samples may be preferable to Fe$_{50}$Pt$_{50}$ for obtaining a large MAE.  In terms of first principles calculations we aim to explore the origin of this observation, in particular, whether it is it a pure effect of composition or whether it is  also related to the chemical order of the sample.\\*

To this end, we perform fully relativistic first-principles calculations by means of the screened Korringa-Kohn-Rostoker (SKKR) method.  As the method is well documented elsewhere in the literature, see e.g.~Refs.~\onlinecite{laszlo1,screening1,screening2,ebert}, here we describe only the features particularly relevant to this work.  We use the local spin density approximation (LSDA) of the density functional theory (DFT) as parameterized by Vosko et al.\cite{voskoCJP80} and treat the potentials within the atomic sphere approximation (ASA).  In line with previous work,\cite{apl-letter} the self-consistent potentials and fields are calculated from scalar-relativistic calculations and the fully relativistic Kohn-Sham-Dirac equation is then solved to derive the MAE of the system. In all calculations, an overall angular-momentum cut-off of $\ell_{\rm max}=3$ was used.\\*

The chemical disorder according to the model as described above was treated in terms of the coherent potential approximation (CPA).\cite{soven,gyorffy} It should be mentioned that the MAE of Fe$_{x}$Co$_{1-x}$ alloys was recently investigated by using the same model of long-range chemical order.\cite{turek,kota}
In the spirit of the magnetic force theorem,  the MAE is evaluated as the difference in the band energy of the system when polarized along the easy axis $(0 0 1)$ and perpendicular to the easy axis, along $(1 0 0)$. As in Ref.~\onlinecite{apl-letter} we estimated the effect of temperature-induced spin-fluctuations by scaling down the MAE by a factor of 0.6 according to the Langevin dynamics simulations of Mryasov {\em et al.}~\cite{mryasov}\\*

\begin{figure}
\centering
\includegraphics[scale=0.33]{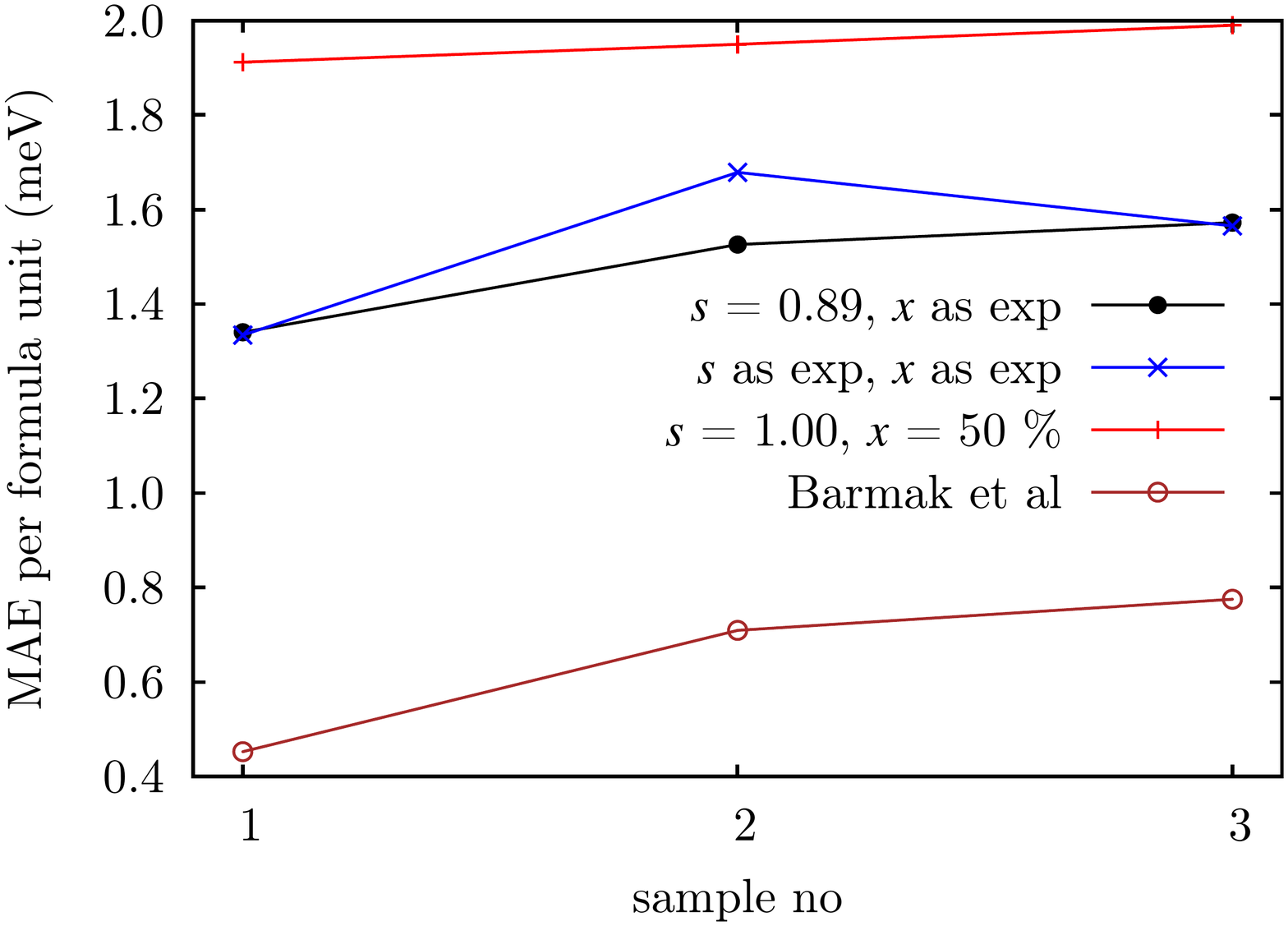}
\caption{A comparison of the experimental MAE values of the Fe$_x$Pt$_{1-x}$ samples studied by Barmak {\em et al.} in Ref.~\onlinecite{barmak} (open circles) and theoretical MAE values calculated using SKKR-CPA as follows,
$+$ ~:  using the experimental lattice parameters for each sample, but assuming $x=0.5$ and $s=1$,   $\bullet$~: using the lattice parameters and the compositions $x$ as given in the experiment, but keeping $s$ constant at 0.89,  and $\times$~: using the lattice parameters  as well as the values of $x$ and $s$ as given in the experiment. Solid lines serve as guide for the eyes.}
\label{fig:barmaksamples}
\end{figure}

In order to verify our method against the experiments, first we attempt a direct comparison of our SKKR-CPA calculations to the experimental data of Ref.~\onlinecite{barmak},  see Table \ref{table:samplesBarmak}.  As shown in Fig.~\ref{fig:barmaksamples}, we performed three sets of calculations.  The first set only takes into account changes in the lattice geometry (i.e., the variation in the lattice parameters), while assuming stoichiometric composition, $x=0.5$, and maximum long-range chemical order, $s=1$.  Even by taking into account the 'temperature factor' of 0.6, these calculations yield quite high MAE values. Such magnitude differences with respect to the experiment are, however, in agreement with previous first-principles calculations of the MAE of FePt, see e.g.~Refs.~\onlinecite{lyubina,shick}. Furthermore, in this set of calculations only a very moderate change ($<3 \%$) of the MAE is obtained across the samples.\\*

In the second set of calculations, we introduce the composition $x$ as given in experiment, while keeping the degree of chemical order constant at $s=0.89$.  This greatly improves the trend of the MAE, however, the relative change of the MAE from sample no. 1 to sample no. 2 is still underestimated ($< 15 \%$)  as compared to the experiment ($\sim 50 \%$). The overall magnitude of the MAE is significantly decreased, but it is still by a factor of 2-2.5 larger than the measured one.  Finally, including also the variation of chemical order as given in the experiment clearly improves the above mentioned relative change between samples no 2 and 3 ($\sim 30 \%$), but, opposite to the experiment, it predicts a slightly decreasing  trend from sample no. 2 to 3.  Note, however, that these latter changes are within the range of both theoretical and experimental errors.  \\*

\begin{figure}[htp!]
\centering
\includegraphics[scale=0.33]{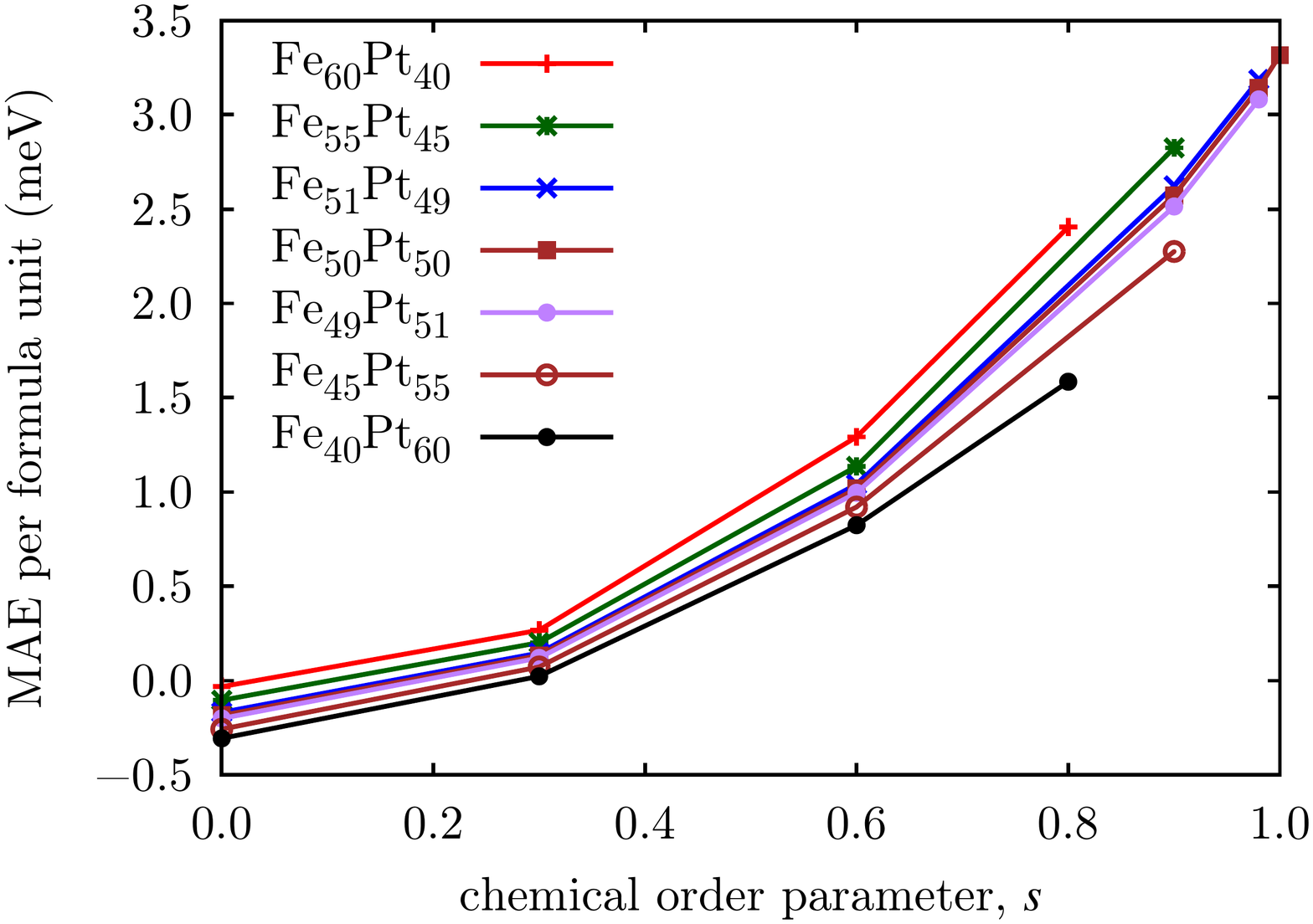}
\caption{The variation of the MAE of Fe$_x$Pt$_{1-x}$ alloys as a function of the chemical order parameter $s$ and the composition $x$. 
Solid lines serve as guide for the eyes.
\label{fig:MAE_xsa}}
\end{figure}

Having confirmed that the SKKR-CPA calculations satisfactorily reproduce the experimental trends, we next consider the general effects of the chemical composition $x$ and order parameter $s$ on the MAE of FePt.  To this end, we used the lattice parameters measured for sample no. 3 in Ref.~\onlinecite{barmak}, $a=3.857$ \AA~and $c=3.706$ \AA, while we independently varied the chemical order parameter $s$ as well as the composition $x$.  Note that for this theoretical study we did not scale down the MAE to mimic temperature induced effects.  The results are shown in Fig.~\ref{fig:MAE_xsa} for the range of compositions $0.4 \leq x \leq 0.6$.  (Beyond this range, the L1$_0$ structure becomes unstable with respect to other phases.\cite{whang, massalski})  Our results are in good agreement with the conclusion of Barmak {\em et al}\cite{barmak}, inasmuch for any given degree of chemical order $s$ the MAE increases monotonically with the Fe-content.  However, even maximally ordered Fe$_x$Pt$_{1-x}$ alloys with $x > 50$ cannot achieve the MAE of fully ordered Fe$_{50}$Pt$_{50}$ (3.31 meV per formula unit).\\*

It should be noted that, at $s=0$, the MAE becomes negative. This is in contrast to Ref.~\onlinecite{ostanin}, which reports a vanishing MAE for completely disordered FePt under the assumption of a cubic unit cell.  The 'residual' negative MAE we obtain in the case of complete chemical disorder is, therefore, due to the tetragonality of the lattice ($a \ne c$).  For real samples, where the lattice parameters cannot be frozen while varying the chemical order and composition, in the case of complete chemical disorder the unit cell is expected to become cubic, removing thus this 'residual' MAE.\\*

\begin{figure}
\centering
\includegraphics[scale=0.6]{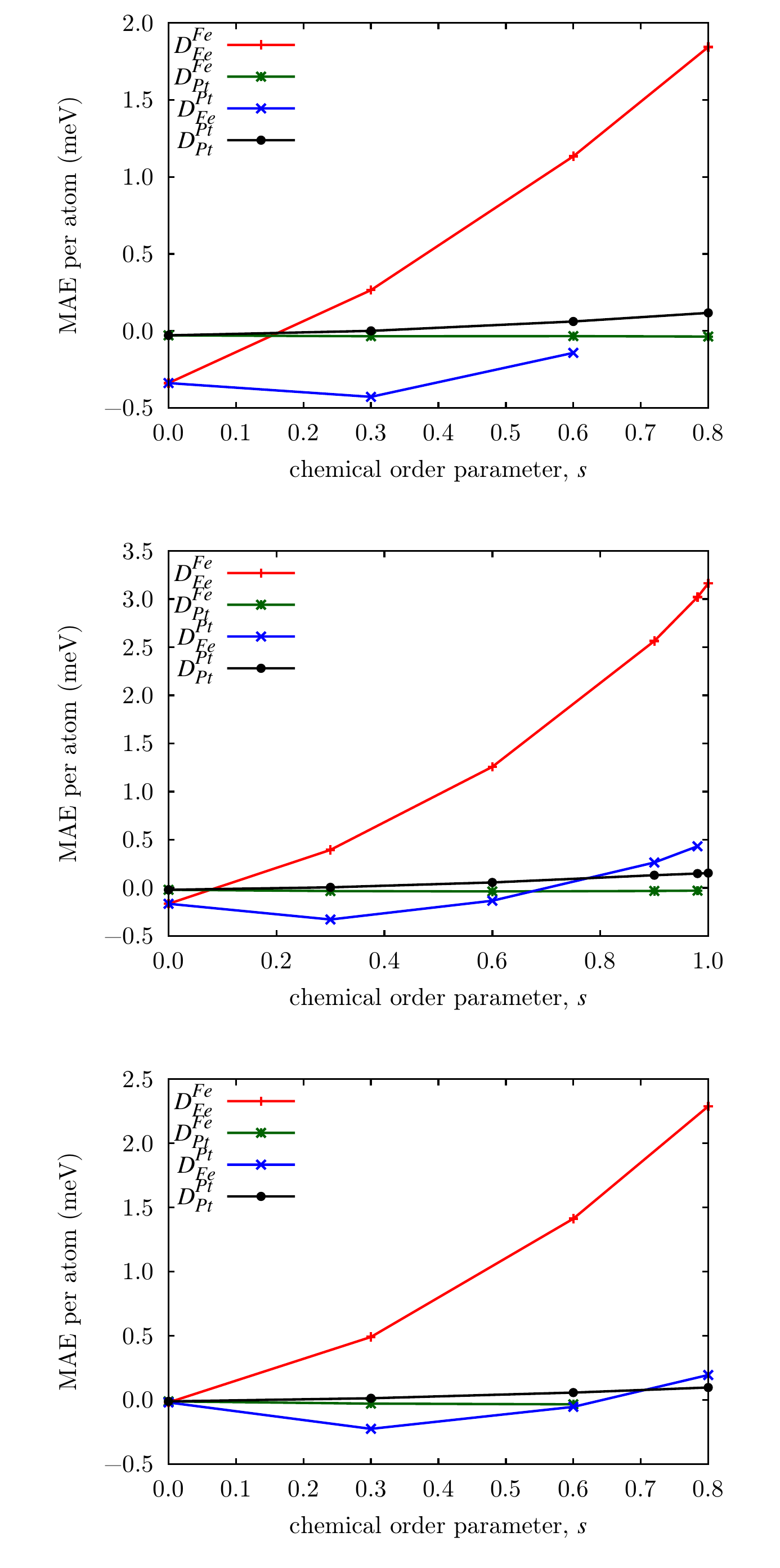}
\caption{Variation in the species-resolved MAE contributions against the chemical order parameter $s$ for compositions
 $x=0.40$ (upper panel), $x=0.50$ (middle panel) and $x=0.60$ (lower panel).   $+$ : $D_{Fe}^{Fe}$, contribution of an Fe atom in nominal Fe layers, $\ast$ : $D_{Pt}^{Fe}$, contribution of a Pt atom in nominal Fe layers, $\times$ : $D_{Fe}^{Pt}$, contribution of an Fe atom in nominal Pt layers,
 and  $\bullet$ :  $D_{Pt}^{Fe}$, contribution of a Pt atom in nominal Pt layers. Solid lines serve as guide for the eyes.
\label{fig:MAElay}}
\end{figure}

In order to elucidate the origin of the variation in the MAE with the composition and the chemical disorder, we consider next the species-resolved contributions to the MAE.  The MAE per unit cell can be decomposed as
\begin{equation}
 K = r_{Fe} D_{Fe}^{Fe} + (1-r_{Fe})D_{Pt}^{Fe} + (1-r_{Pt}) D_{Fe}^{Pt} + r_{Pt}D_{Pt}^{Pt} \; ,
 \label{eq:Kcontr}
\end{equation}
where $D_{\gamma}^{\beta}$ ($\beta, \gamma =$ Fe or Pt) denotes the MAE contribution from an atom of species $\gamma$ when it is positioned in a nominal $\beta$ layer, i.e., within a layer, which in a perfectly ordered Fe$_{50}$Pt$_{50}$ alloy would contain only atoms of species $\beta$.  For the cases of $x=0.4$, 0.5 and 0.6, in Fig.~\ref{fig:MAElay} we show $D_{\gamma}^{\beta}$ as a function of the chemical order parameter, $s$. In completely disordered FePt ($s=0$), the nominal Fe layers and the nominal Pt layers are identical.  Therefore, at $s=0$, the Fe contributions in both layers are equal and take a small negative value for $s=0$, which decreases in magnitude with increasing $x$ and practically vanishes at $x=0.6$.  The Pt contributions, on the other hand, are nearly zero for all compositions $x$ when $s=0$.  As the chemical order $s$ increases, the Fe contribution in the nominal Fe layers rapidly increases up to about 1.8 meV, 2.0 meV and 2.2 meV  at $s=0.8$ for $x=0.4$, 0.5 and 0.5, respectively. For the fully ordered case, $x=0.5$ and $s=1$, $D_{Fe}^{Fe}$ even takes the value of about 3.15 meV, close to the total value of the MAE (3.31 meV). In contrast, the Fe contribution in nominal Pt layers decreases up to $s \simeq 0.3$, then slightly increases and, for $x \ge 0.5$, reaches a small positive value ($< 0.5$ meV) at maximal chemical order. Remarkably, the magnitude of the Pt contributions remain almost negligible ($< 0.15$ meV) over the whole range of chemical order.\\*

\begin{figure}[htp!]
\centering
\includegraphics[scale=0.33]{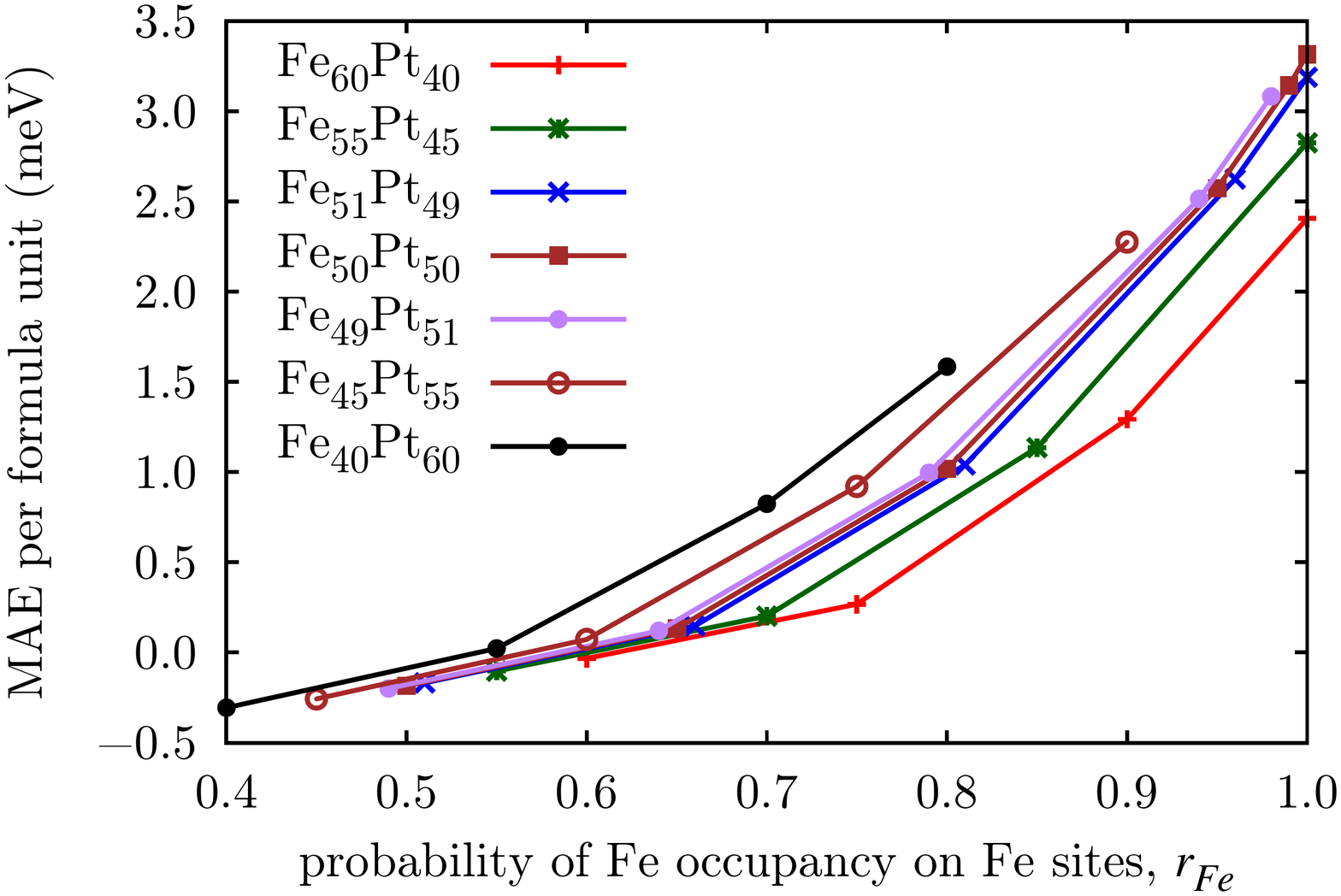}
\caption{The variation of the MAE of Fe$_x$Pt$_{1-x}$ alloys as a function of the Fe concentration in the nominal Fe layers $r_{Fe}$ and the overall Fe concentration $x$. Solid lines serve as guide for the eyes. \label{fig:MAE_xsb}}
\end{figure}

As indicated in Fig.~\ref{fig:MAElay}, the dominant contribution to the MAE is $r_{Fe} D_{Fe}^{Fe}$, see Eq.~(\ref{eq:Kcontr}). It is, therefore, intuitive to replot Fig.~\ref{fig:MAE_xsa} as a function of the fraction, $r_{Fe}$. This interpretation of the MAE is shown in Fig.~\ref{fig:MAE_xsb} for different compositions $x$. Since $r_{Fe}=x+\frac{s}{2}$, it is clear that the horizontal range of the curves in Fig.~\ref{fig:MAE_xsa} is halved and, more importantly, they are shifted to the right by $x$. As a consequence, for a fixed value of $r_{Fe}$  the order of the curves with respect to $x$ is reversed as compared the order of curves at a given $s$ in Fig.~\ref{fig:MAE_xsa} . This opposite tendency becomes obvious when considering e.g. the case of $r_{Fe}=1$ -- Fe-rich Fe$_x$Pt$_{1-x}$ will exhibit completely Fe-filled nominal Fe layers at a smaller $s$ than Fe$_{50}$Pt$_{50}$, which requires $s=1$ in order to exhibit completely filled Fe layers. On the other hand,  increasing disorder (i.e., decreasing $s$) drastically reduces the Fe contribution to the MAE in the nominal Fe layer, $D_{Fe}^{Fe}$, and, consequently, the MAE of the system. \\*

In conclusion, our calculations strongly support the conclusion of Barmak and co-workers in Ref.~\onlinecite{barmak}, showing that, for a given degree of chemical order, the MAE increases with the Fe concentration of Fe$_x$Pt$_{1-x}$, at least within the range $0.4 \leq x \leq 0.6$.  This is due to the strongly positive effect on the MAE of the degree of Fe-filling of the nominal Fe layers, $r_{Fe}$, which dominates the variation in the MAE when varying the composition $x$, while keeping the chemical chemical disorder $s$ constant.  However, Fe$_{x}$Pt$_{1-x}$ with $x\neq 0.50$ cannot attain perfect chemical order ($s=1$) and perfectly ordered Fe$_{50}$Pt$_{50}$ yields a larger MAE than the Fe-rich alloys with maximum degree of long-range chemical order.\\*

The authors would like to thank and acknowledge Professor Katayun Barmak and Professor Jingsheng Chen for helpful discussions and advice on the details of the experimental work in Refs.~\onlinecite{expl,barmak}. CJA is grateful to EPSRC and to Seagate Technology for the provision of a research studentship.
Support of the EU under FP7 contract NMP3-SL-2012-281043 FEMTOSPIN is gratefully acknowledged.
Financial support was in part provided by the New Sz\'echenyi Plan of Hungary (T\'AMOP-4.2.2.B-10/1--2010-0009)
and the Hungarian Scientific Research Fund (OTKA K77771). \\*

\end{document}